\documentclass[referee]{raa}            

\usepackage{graphicx,times}             

\begin{document}

   \title{The peculiar velocity and temperature profile of galaxy clusters
}

   \volnopage{Vol.0 (200x) No.0, 000--000}      
   \setcounter{page}{1}          

   \author{Tabasum Masood
      \inst{1}
   \and Naseer Iqbal
      \inst{1,2}
	}

    \institute{ Department of Physics, Kashmir University, Srinagar 190 006, 
India; {\it iqbal@iucaa.ernet.in}\\
        \and
             Inter University Centre for Astronomy and Astrophysics, IUCAA, Pune, 411007, India\\
     }

   \date{Received~~2009 month day; accepted~~2009~~month day}

\abstract{ Dynamical parameters like average velocity dispersion and temperature profile of galaxy clusters are determined using a quasi equilibrium thermodynamic theory. The calculated velocity  dispersion results from theory and simulations shows a good agreement with the velocity  dispersion results of (Abdullah et al.~\cite{Abd11}). An Adaptive Mesh Refinement (AMR), Grid based hybrid code have been used to carry out the simulations. Our results indicate that the average velocity  dispersion profile of $20$ Abell galaxy clusters falls in the range of $500-1000$ km/s  and their temperature profile is of the order of $10^7$ to $10^8$K calculated on the basis of kinetic theory. The data in the plot shows a significant contribution of gravitating particles clustering together in the vicinity of cluster center and beyond a certain region this velocity dies out and gets dominated by the Hubble's flow due to which all the galaxy clusters in an expanding universe participate in Hubble's expansion.
}

   \authorrunning{Tabasum \& Naseer }            
   \titlerunning{ The peculiar velocity and temperature profile of galaxy clusters}  

   \maketitle

%
%
\section{Introduction}           
\label{sect:intro}

In recent years, the developments in the study of large scale structure formation in an expanding universe has become one of the exciting problems in cosmology. The auto-correlation functions, rapid computer simulations, theoretical analysis and observational evidences have added a considerable knowledge in this study. Earlier studies on the applicability of statistical mechanics and thermodynamics to the cosmological many body problem have yielded many interesting results and  have also explored many interesting questions (Saslaw \& Hamilton.~\cite{Sas84}, Saslaw \& Fang.~\cite{Sas96}, Ahmad et al.~\cite{Ahm02}, Iqbal et al.~\cite{Iqb06}, Iqbal et al.~\cite{Iqb11} and Iqbal et al.~\cite{Iqb12}). The time scale for cluster formation is much larger than the time scale of a galaxy to cross the cluster. Although  a cluster still evolves, the galaxies in the cluster form a quasi-equilibrium distribution. Such distributions and evolution are governed by gravitational instability (Saslaw.~\cite{Sas99}). One of the interesting unknowns in the study of large scale structure formation in an expanding universe is peculiar motions which can be studied by using galaxies or cluster of galaxies. These motions provide an important tool in probing the peculiar velocity field. The peculiar velocity of galaxy clusters which represent their departures from the global cosmic expansion provides a significant information about the related details of the cosmological many body problem and the existence of dark matter component. The use of the virial theorem $\textless2T\textgreater+\textless W\textgreater$=$0$ relating the time averages of its kinetic and potential energies is even studied for the effects of configuration on the peculiar motion of galaxy clusters. Some studies have already been carried out for the studying the spatial configuration of galaxy clusters on the basis of virial theorem (Saslaw.~\cite{Sas87}, Saslaw.~\cite{Sas00} and Barychev et al.~\cite{Bar01}). The first peculiar velocity of galaxy clusters was reported by (Gudehus.~\cite{Gud73}). Some earlier work also suggests that these peculiar velocities were referenced to the MBR (Gudehus.~\cite{Gud78}) and the accurate measurements were reported by (Smoot et al.~\cite{Smo77}).

      In this work, we study peculiar velocity of galaxy clusters by using kinetic equation of state and assuming that the system is virialized in which the growth of peculiar motions of various gravitating particles lead to the study of cluster formation. We describe N-body simulations using ENZO $2.1$ hydrodynamical code (Oshea et al.~\cite{Osh05})  which is an Adaptive Mesh Refinement(AMR) code. The peculiar velocity results generated from the simulations are analyzed in detail to look for their comparison  with the results obtained from theoretical calculations based on quasi equilibrium thermodynamic approach. The more confirmations in this context is to see the overall matching of  velocity  dispersion results generated from kinetic theory and simulations with the results of velocity  dispersion profile of (Abdullah et al.~\cite{Abd11}).\\
It is also interesting to examine the temperature profile of galaxy clusters on the basis of same theory with an assumption that galaxy clusters are constituents of hot gas particles which contribute to the overall temperature for a galaxy cluster. We also caliberate  intra cluster temperature for a group of Abell galaxy clusters. The overall impression in this work is to come up with a simple and fundamental approach of studying various dynamical parameters like peculiar motions and temperature for $20$ Abell galaxy clusters in an expanding universe. In \S $2$ we study the basic theoretical formalism of peculiar velocities followed by N-body simulations. \S $3$ describes the intracluster temperature. Finally in \S $4$ we discuss the results.

\section{Peculiar Motion of Galaxy Clusters}
\label{sect:Motion}
 Peculiar motion studies shows that there is a significant gas motion in galaxy clusters with velocities upto $500-1000$ km/s. The description of  cluster peculiar velocity  field is an important step towards the understanding of the nature of the forces acting on the galaxy clusters. This peculiar velocity field can be studied by using galaxy clusters  (Bahcall et al.~\cite{Bah94}, Lauer \& Postman~\cite{Lau94}, Bahcall \& Oh~ \cite{Bah96}, Moscardini et al.~\cite{Mos96}, Borgani et al.~\cite{Bor97}, Borgani et al.~\cite{Bor00}, Watkins~\cite{Wat97}, Dale et al.~\cite{Dal99} and Abdullah et al.~\cite{Abd11}). The observational studies (Rines \& Diaferio~\cite{Rin06}), N-body simulations (Wojtak et al.~\cite{Woj05}, Mamon et al.~\cite{Mam04}, and Cuesta et al.~\cite{Cue08}) and a combination of both (Mahajan et al.~\cite{Mah11}) have shown that the virialized clusters in gravitating systems are sorrounded by infall zones from which most of the galaxies move into the relaxed cluster as is evident from the earlier result of (Gunn \& Gott.~\cite{Gun72}). In order to understand the simple formalism of peculiar velocity, we make use of quasi equilibrium thermodynamics in which quasi equilibrium evolution takes place through a sequence of quasi equilibrium states whose properties change with time (Saslaw \& Hamilton.~\cite{Sas84}, Saslaw et al.~\cite{Sas90}, Ahmad et al.~\cite{Ahm02}, Iqbal et al.~\cite{Iqb06} and Iqbal et al.~\cite{Iqb12}).\\
To determine peculiar motion of galaxy clusters with fixed number of galaxies in a spherical volume of cluster size $R$, the average kinetic energy of peculiar motions of $N$ galaxies in a cluster can reasonably be connected with average kinetic temperature $T$ of a cluster (Saslaw.~\cite{Sas86}) defined as

\begin{equation}
  K~=~ \frac{3}{2}NT ~=~\frac{1}{2}MV^{2}
  \label{eq:1}
\end{equation}

Here $M$ is the mass of a cluster containing $N$ galaxies and $V$  represents the average velocity dispersion of a galaxy cluster. The average  peculiar velocity dispersion of a sample of galaxy clusters can be determined by approximating the dynamical time scale $t_{dyn}$ and crossing time scale $t_c$ of a relaxed cluster of galaxies. The condition $t_{dyn} \ \approx \ t_c$ is valid for a system to be in a state of quasi-equilibrium where the system can be described as virialized and the required condition for a cluster to be in virial equilibrium is that the macroscopic evolution of a system in quasi-equilibrium is slow as compared with the crossing time of the system, so that equilibrium prevails approximately. The description of the velocity dispersion can be understood in knowing its dependence on the respective time scales and a suitable choice for the time unit is a representation of the dynamical scale for the cluster (Yang \& Saslaw~ \cite{Yan11})
\begin{equation}
 t_{dyn}~=~ \frac{1}{\sqrt {G\rho}}~=~\sqrt{\frac{R^3}{GM}}
  \label{eq:2}
\end{equation}
where $\rho$ is the density of a given cluster, $R$ the size of a cluster and $G$ the gravitational constant
\begin{equation}
  V_{pec}~=~\sqrt{\frac{3GM}{\pi R}}
  \label{eq:3}
\end{equation}
This is average velocity dispersion of a cluster.\\
Table$~1$ shows the  velocity  dispersion profile of a sample of $20$ Abell galaxy clusters using $SDSS-DR7$ at virial radius $1-2h^{-1}Mpc$ (Abdullah et al.~\cite{Abd11}). The calculations shown in column $5$ are average peculiar velocity dispersion results calculated from equation $3$. While comparing the two sets of results shown in column $4$ and $5$, the comparison is in good agreement.
\subsection{THE SIMULATIONS}
We describe here $N$ body simulations for understanding the peculiar motions of galaxy clustering in an expanding universe. The $N$ body simulation have been carried out by using ENZO $2.1$ hydrodynamical code (Oshea et al.~\cite{Osh05}). $ENZO$
v.$2.1$ is an Adaptive Mesh Refinement(AMR), grid-based hybrid (N-body plus hydrodynamical) code. We have used a flat LCDM (Lambda Cold Dark Matter) background cosmology with the parameters defined by $\Omega_{mo}=0.27$, $\Omega_{\Lambda o}= 0.723$, $\Omega_{b}=0.0459 $, $n_s=0.962$, $H_0=70.2km/s/Mpc$ and $\sigma_8=0.817$  derived from WMAP 5 (Komatsu et al.~\cite{Kom09}). The simulations have been initialized at redshift $z=60$ using the (Eisenstein \& Hu.~\cite{Eis99}) transfer function, and evolved upto $z=0$. An ideal equation of state was used for the gas, with $\gamma
=\frac{5}{3}$. Radiative cooling assumed from (Sarazin.~\cite{Sar87})  for a fully ionized gas with metallicity of $0.5$ solar.

The simulation was performed in a box of comoving volume $(64$
${h^{-1} \ \rm Mpc})^3$ containing $64^3$ particles, each having $9.05 \times 10^{10} \ M_{\odot}$, with 3 levels of AMR. This gives the effective highest resolution of $128$ kpc.
The halo centers have been identified using enzohop algorithm
(Eisenstein \& Hut.~\cite{1998ApJ...498..137E}). Considering a particular center obtained from enzohop, we calculate the density within a spheriodal shell. The shell radius is increased until it reaches the virial radius. The virial radius is defined as that radius where the average density of the spheroidal shell equals to the $200$ times of critical density. We calculate the peculiar velocity as the RMS velocity of the galaxy clusters inside the virial radius.  For estimating the cosmic variance, the analysis was carried out for $10$ different independent realization of the simulation.

For comparision, the variation of velocity  dispersion with virial cluster radius for three different data sets shown in Figure$.1$ clearly shows that the theory and the simulations results are in good agreement with the (Abdullah et al.~\cite{Abd11}).

\section{Kinetic Cluster Temperature}

The kinetic cluster temperature of galaxy clusters  is defined in terms of the average kinetic energy of its constituent particles (galaxies) and can be described by a quasi equilibrium thermodynamic system. It has been known for years that baryons in a DM dominated potential, will heat up to a temperature that is determined by the properties of DM mass profile (Rees \& Ostriker.~\cite{Ree77}, Cavaliere \& Fusco.~\cite{Cav78}, Cavaliere \& Fusco.~\cite{Cav81}, Sarazin.~\cite{Sar86} and Cavaliere et al.~\cite{Cav09}). In hydrostatic equilibrium, the total mass within a given radius is proportional to the local temperature at that radius. Thus it is imperative to measure the cluster temperature at large radii to a good precision. The largest  samples to date used to measure temperature profiles, with ASCA (Markevitch et al.~\cite{Mar98}), BeppoSAX (De Grandi \& Molendi.~\cite{De02}), XMM-Newton (Pointecouteau et al.~\cite{Poi05}), and Chandra (Vikhlinin et al.~\cite{Vik05}) data. The deep observational study, the swift XRT able to measure the ICM profiles in the external region of virialized radii has shown the best results compared with the previous measurements (Moretti et al.~\cite{Mor11})\\
Intra cluster gas, the hottest thermal equilibrium plasma has been heated to temperature of tens of millions of degrees (kelvin) which cause hot gas to emit the bulk of thermal energy range (Bohringer \& Werner.~\cite{Boh10}).  Here  our velocity dispersion profiles allow us to estimate the temperature of  galaxy cluster with an assumption that cluster of galaxies are homologous and intracluster gas is in global equilibrium with the dark matter. In this case, where this gas is an ideal gas, temperature of hot gas is set up by virial condition
\begin{equation}
\frac{3}{2}Nk_{B}T_{cl}~=~ \frac{1}{2}\sum_{i=1}^{N} m_{i}V_i^2
\end{equation}
where $m_i$ are masses of constituent particles and $k_B$ is Boltzmann constant. The calculation of cluster gas temperature at a location is given by local velocity distribution. Here this velocity is of the order of average velocity  dispersion of a cluster. Considering this velocity as the order of magnitude of a mean velocity $(V_{mean})$ of the hot gas molecules, the above equation can be written as
\begin{equation}
\frac{3}{2}NK_{B}T_{cl}~=~ \frac{1}{2} m V_{mean}^2
\end{equation}
where $m$ is the mean molecular weight which is approximate atomic weight for hydrogen atom. We calculate here the cluster temperature from above equation using peculiar velocity values shown in column $5$ in Table$~1$. The values of $m$ and $V$ predict temperature in the range of millions of degrees of kelvin. So we expect cluster to be filled with an X-ray emitting plasma.

\begin{figure*}
  \centering 
  \rotatebox{0}{\scalebox{.6}{\includegraphics{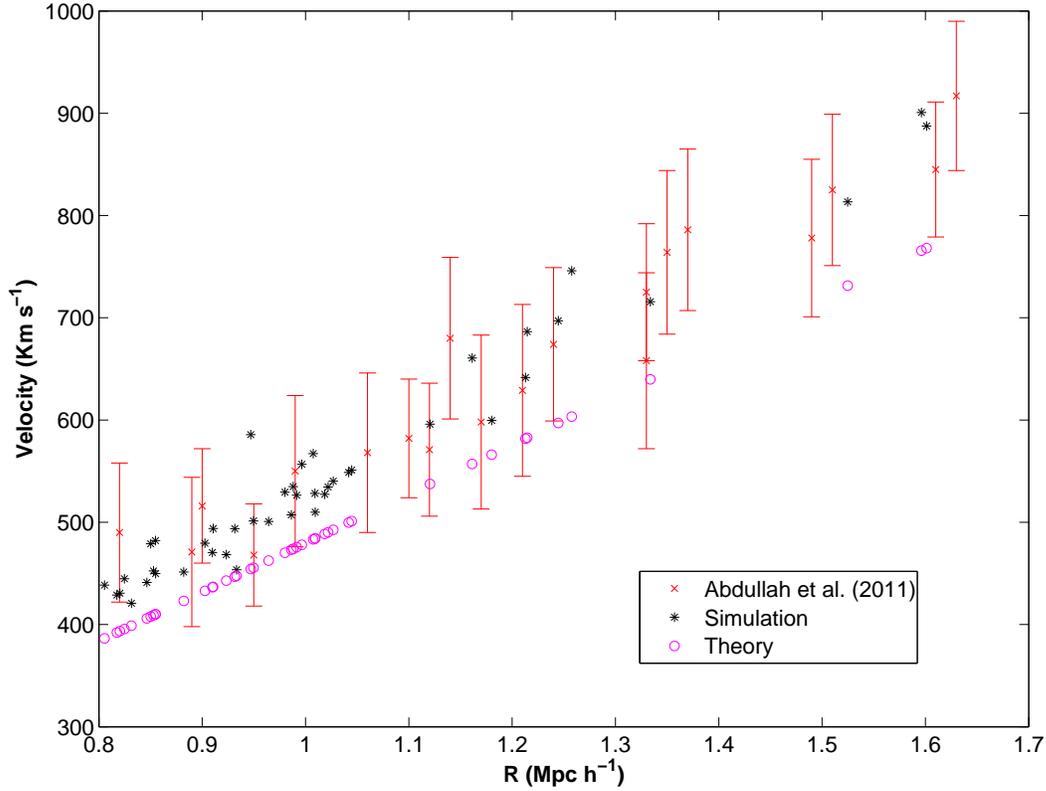}}}
  \caption{Velocity variation vs Radius for three different data
    sets.}
  \label{fig:1}
\end{figure*}
\begin{table*}

\caption{Peculiar velocity distribution of galaxy cluster samples  studied at virial radii $(1-2h^{-1}Mpc)$}
\begin{center}
\begin{tabular} {l c c c c }
\hline  
Name & R & Cluster Mass  & $V_{pec} km/s$ & $V_{pec}$ \\ 
     & $(1-2h^{-1} Mpc)$& $(10^{14}M_{\odot})$ & (Abdullah et al.~\cite{Abd11}) &$(km/s)$ \\

\hline
A0117 & 1.216 & 2.7 & $565\pm63$& 600\\
A0168 & 1.204 & 2.1 & $576\pm56$& 615\\
A0671 & 1.434 & 4.9 & $786\pm79$& 830\\
A0779 & 1.065 & 1.61& $519\pm52$& 550\\
A1066 & 1.446 & 4.7 & $750\pm77$& 804\\
A1142 & 1.093 & 3.3 & $540\pm72$& 590\\
A1205 & 1.239 & 5.1 & $603\pm84$& 660\\
A1238 & 1.022 & 2.9 & $477\pm70$& 510\\ 
A1377 & 1.011 & 2.1 & $498\pm61$& 532\\ 
A1424 & 1.154 & 3.7 & $562\pm75$& 614\\
A1436 & 1.379 & 3.6 & $662\pm69$& 700\\
A1459 & 0.96  & 1.5 & $510\pm54$& 538\\
A1663 & 1.261 & 4.7 & $618\pm81$& 668\\
A1767 & 1.634 & 4.8 & $827\pm74$& 874\\
A1809 & 1.247 & 4.0 & $674\pm76$& 719\\
A2048 & 1.387 & 5.7 & $652\pm85$& 707\\
A2061 & 1.434 & 3.4 & $722\pm67$& 756\\
A2142 & 1.809 & 3.5 & $844\pm62$& 860\\
A2255 & 1.757 & 4.5 & $904\pm70$& 937\\
A2670 & 1.557 & 4.7 & $770\pm75$& 813\\

\hline
\end{tabular}
\end{center}
\end{table*}

\section{Discussion}
\label{sect:discussion}
In this work, we present a quasi-equilibrium thermodynamic approach for studying the velocity  dispersion profile and temperature profile of galaxy clusters in an expanding universe. Peculiar velocity field provides a basic understanding  of the clustering rate of  galaxies. The calculated velocity  dispersion results obtained from equation$.3$ fall in the range of $500-1000$ km/s for rich $20$ Abell galaxy clusters and therefore provides a good approximation that the virialized systems of galaxy clusters can be well defined by the equation of state. An attempt has been made to carry out N-body simulations for the calculation of peculiar velocity field by using ENZO $2.1$ hydrodynamical code.The simulation results and peculiar velocity results obtained by quasi equilibrium approach shows a good agreement with (Abdullah et al .~\cite{Abd11}). The growth of clustering on the basis of peculiar velocity field is illustrated in Figure$~1$. This is a plot of average velocity  dispersion of each sample of galaxy cluster with cluster size. For smaller values of cluster size, the number density is large which corresponds to higher rate of clustering and with the increase in size from cluster centre to outer regions, the clustering rate weakens due to which global Hubble's flow becomes dominant.\\
The temperature profile of $20$ Abell galaxy clusters making use of kinetic equation comes out to be of the order of cluster temperature of hot X-ray gas of clusters.

\normalem
\begin{acknowledgements}
The authors are thankful to Prakash Sarkar and Surajit Paul Post Doc fellows of IUCAA, Pune, India for helping in carrying out N-body simulations in this work. The authors are also highly grateful to Inter University centre for Astronomy and Astrophysics (IUCAA), Pune, India for their hospitality during our stay.

\end{acknowledgements}

\label{lastpage}

\end{document}